\documentclass{icrc}
\usepackage{amssymb,amsmath}
\usepackage{times}
\newif\ifpdf 
\ifx\pdfoutput\undefined
\pdffalse    
\else
\pdfoutput=1 
\pdftrue
\fi \ifpdf   
    \usepackage[pdftex]{graphicx} 
\else
    \usepackage{graphicx}         
\fi
\begin{document}

\title{Atmospheric proton and neutron spectra at energies above 1 GeV}
\author[1,2]{V. A. Naumov}
\affil[1]{Dipartimento di Fisica and Sezione INFN di Ferrara,
             Via del Paradiso 12, I-44100 Ferrara, Italy}
\author[2]{T. S. Sinegovskaya}
\affil[2]{Laboratory for Theoretical Physics, Irkutsk State
          University, Gagarin boulevard 20, RU-664003 Irkutsk,
          Russia}

\ifpdf
\correspondence{T.\,S.\,Sinegovskaya (tanya@api.isu.run net.ru)}
\else 
\correspondence{T. S. Sinegovskaya (tanya@api.isu.\\runnet.ru)}
\fi

\runninghead{V.\,A.\,Naumov and T.\,S.\,Sinegovskaya: Atmospheric
             proton and neutron spectra at energies above 1 GeV}
\firstpage{1}
\pubyear{2001}
\maketitle

\begin{abstract}
We discuss an effective numerical method for solving the transport
equations for cosmic ray nucleons in the atmosphere. It is
demonstrated that the nucleon attenuation lengths are strongly energy
and depth dependent due to the non-power-law behavior of the primary
spectrum, growth of the total inelastic cross sections with energy,
and scaling violation in the nucleon-nucleus interactions. The
numerical results are compared with the available experimental data.
\end{abstract}

\protect\section{Introduction} \label{sec:Introduction}

Measurements of the fluxes of secondary cosmic-ray protons and
neutrons can furnish valuable information about primary cosmic rays
and about the nuclear interactions at high energies. In order to
extract this information from experimental data, it is necessary,
among other things, to be able to calculate the nucleon attenuation
lengths, which are functionals of the primary spectrum and which also
depend, in general, on energy and atmospheric depth.

In this paper we discuss some results obtained by using a simple
but rather efficient and physically evident numerical method for
solving transport
equations describing the propagation of cosmic-ray protons and
neutrons through the atmosphere \citep{Naumov00}. The method is
applicable at sufficiently high energies and demands several
approximations which are quite traditional for high-energy
atmospheric cascade calculations. Namely, we use the
one-dimensional (1D) approach, in which all secondary particles in
the cascade are supposed collinear with the projectiles. The
processes of generation of $N\overline{N}$ pairs in meson-nucleus
collisions are disregarded, but the corresponding contribution is
typically small and can be taken into account as a correction
\citep{Vall86}. Besides, we apply the standard superposition model
for the collisions of cosmic-ray nuclei and neglect the
geomagnetic effects and proton ionization energy losses. These
approximations confine the range of applicability of the method.
However, it remains rather broad in order to describe all
currently available data on the high-energy nucleons in the
atmosphere and at sea level. In particular, it covers completely
the depth-energy area relevant to calculations of atmospheric muon
and neutrino fluxes at energies above several GeV, -- one of the
most important fields of application of the method.

We compare our calculations with the experimental data and with
the results of a more sophisticated approach by
\citet{Fiorentini01} based on an updated version of CORT code
which was designed for applications to low and intermediate energy
atmospheric cascade calculations and which takes into account
essentially all significant effects
\citep{Naumov84,Bugaev85a,Bugaev85b,Bugaev98}.

\protect\section{The {\em Z} factor method}

Within the above assumptions, the problem of calculating the
differential energy spectra of protons $D_p(E,h)$ and neutrons
$D_n(E,h)$ at the depth $h$ consists in solving the following set
of 1D transport equations:
\begin{align}\label{TranspEq}
 \left[\frac{\partial}{\partial h}\right.
&+\left.\frac{1}{\lambda_N(E)}\right]D_N(E,h)=
 \sum_{N'}\frac{1}{\lambda_{N'}(E)}\int^\infty_EdE' \nonumber\\
&\times\frac{1}{\sigma_{N'A}^{\mathrm{inel}}(E)}
 \frac{d\sigma_{N'N}(E',E)}{dE}D_{N'}(E',h)
\end{align}
($N,N'=p,n$), with the boundary conditions
\begin{equation} \label{BoundCond}
D_p(E,0)=D_p^0(E), \quad D_n(E,0)=D_n^0(E),
\end{equation}
where $D_p^0(E)$ and $D_n^0(E)$ are the differential energy spectra
of protons and neutrons at the top of the atmosphere;
$\lambda_N(E)=1/[N_0\sigma_{NA}^{\mathrm{inel}}(E)]$ is the
nucleon interaction length;
$d\sigma_{N'N}(E',E)/dE$ is the differential cross section for
inclusive reaction $N'A \to NX$; $E'$ and $E$ are the total
energies of the projectile and final nucleons, respectively.

The approximate isotopic symmetry of $NA$ interactions makes it
possible to reduce the set of equations (\ref{TranspEq}) to two
independent equations for the linear combinations
\begin{equation}\label{N_pm}
N_\pm(E,h)=D_p(E,h) \pm D_n(E,h).
\end{equation}
Let us write these combinations in the form
\begin{equation}\label{Ansatz}
N_\pm(E,h)=N_\pm(E,0)\exp\left[-\frac{h}{\Lambda_\pm(E,h)}\right].
\end{equation}
Below, the functions $\Lambda_\pm(E,h)$ will be referred to as
effective attenuation lengths. It is also convenient to introduce
the dimensionless functions
\begin{equation}\label{Z}
Z_\pm(E,h)=1-\lambda_N(E)/\Lambda_\pm(E,h)
\end{equation}
(``$Z$ factors''). In just the same way as the effective
attenuation lengths, the $Z$ factors contain full information about
the kinetics of nucleons in the atmosphere. Substituting Eqs.
(\ref{N_pm}), (\ref{Ansatz}), and (\ref{Z}) into transport
equations (\ref{TranspEq}) and integrating by part, we find that
the $Z$ factors satisfy the integral equation
\begin{align}\label{Zint}
Z_\pm(E,h)&=\frac{1}{h}\int_0^hdh'\int_0^1dx\,\eta_\pm(x,E)
            \Phi_\pm(x,E)  \nonumber\\
          &\times\exp\left[-h'D_\pm(x,E,h')\right],
\end{align}
where
\begin{align*}
D_\pm(x,E,h) &=\frac{1-Z_\pm(E',h)}{\mathstrut\lambda_N(E')}
             -\frac{1-Z_\pm(E ,h)}{\lambda_N(E )}, \\
\Phi_\pm(x,E)&=
\frac{E}{\sigma_{NA}^{\mathrm{inel}}(E)}
\left[\frac{d\sigma_{pp}(E',E)}{dE}\pm
      \frac{d\sigma_{pn}(E',E)}{dE}\right], \\
\eta_\pm(x,E)&=\frac{1}{x^2}\left[\frac{D_p^0(E') \pm D_n^0(E')}
                    {D_p^0(E) \pm D_n^0(E)}\right],
               \quad x=\frac{E}{E'}.
\end{align*}
Although Eq. (\ref{Zint}) is nonlinear, it is much more
convenient to solve it by an iterative procedure than the original
transport equations.%
\footnote{The main reason is in the comparatively weak energy
          dependence of the $Z$ factors (see Fig. \ref{fig:Z}
          below).}
The rate of convergence of the procedure depends on the choice of
zero-order approximation. The simplest choice is
$Z_\pm^{(0)}(E,h)=0$, in which case $D_\pm^{(0)}(x,E,h)$ is
independent of $h$ and in first approximation we have
\begin{align*}
Z_\pm^{(1)}(E,h)=&\int_0^1dx\left[\frac{\eta_\pm(x,E)\Phi_\pm(x,E)}
                  {h/\lambda_N(E/x)-h/\lambda_N(E)}\right]\nonumber\\
                 &\times\left\{1-\exp
              \left[h/\lambda_N(E/x)-h/\lambda_N(E)\right]\right\}.
\end{align*}
Obviously, the recurrent relations for $n$-th approximation (for
$n=1,2,\ldots$) are given by
\begin{align*}
Z_\pm^{(n)}(E,h)&=\frac{1}{h}\int_0^hdh'\int_0^1dx\,\eta_\pm(x,E)
                  \Phi_\pm(x,E)  \nonumber\\
                &\times\exp\left[-h'D_\pm^{(n-1)}(x,E,h')\right], \\
D_\pm^{(n)}(x,E,h)&=\frac{1-Z_\pm^{(n)}(E/x,h)}{\lambda_N(E/x)}
                  - \frac{1-Z_\pm^{(n)}(E  ,h)}{\lambda_N(E  )}.
\end{align*}
Numerical analysis has revealed that the rate of convergence of this
algorithm is quite sufficient for practical uses.

\protect\section{Numerical results and discussion}

In the present calculations we employ three models for the primary
cosmic ray spectrum and composition: NSU
\citep{Nikolsky84,Nikolsky87}, EKS \citep{Erlykin87}, and FNV
\citep{Fiorentini01}. The NSU and EKS models describe the high-energy
range of the spectrum ($\sim10^2$ to $\sim10^8$ GeV/nucleon) and take
into account the change of the spectral index in the ``knee'' region.
The FNV model is a parametrization of the data from the most recent
measurements of the primary spectrum below the ``knee'' ($0.1$ to
about $10^4$ GeV/nucleon) for minimum of solar activity. All three
spectra are extrapolated up to energy $E=E_c=3\times10^{10}$
GeV/nucleon above which a soft cutoff is supposed.

For the differential cross sections $d\sigma_{N'N}(E',E)/dE$ we
use slightly revised semiempirical model by
\citet{Kimel'74,Kimel'75}. For the total inelastic cross section
$\sigma_{pA}^{\mathrm{inel}}$ we apply the parametrization by
\citet{Mielke94}.

Our calculations based on the $Z$ factor method are performed for the
energy range between 1 and $3\times10^{10}$ GeV at $h\leq4\times10^3$
g/cm${}^2$. It should be noted that, at $h\sim10^3$ g/cm${}^2$, the
energy losses by protons are important up to $E\sim30$ GeV.%
\footnote{By way of example, we indicate that, at $E=10$ GeV,
          the relevant correction is about 20\% for protons and
          about 6\% for neutrons \citep{Bugaev85a,Bugaev85b}.}
The $Z$ factor method (which neglects the energy loss effect) is
extrapolated to the low-energy region in order to match the results
of a more accurate analysis performed with CORT code. In the absence
of experimental data on the fluxes of secondary nucleons arriving
from oblique directions, our calculations for $h>10^3$ g/cm${}^2$
are used at present only to test convergence of the iterative
algorithm.

At all values of $E$ and $h$ iterative process converges fast:  five
to six iterations are sufficient for calculating the $Z$ factors to
precision not poorer than $10^{-3}-10^{-4}$. At moderate depths,
$h\lesssim300$ g/cm${}^2$, even the first approximation ensures a
precision of a few percent, which is sufficient for many applications
of the theory -- in particular, for calculating the fluxes of
atmospheric muons and neutrinos.

\begin{figure}[t]
\vspace*{2.0mm} 
\ifpdf\includegraphics[width=8.3cm]{z.pdf} 
\else \includegraphics[width=8.3cm]{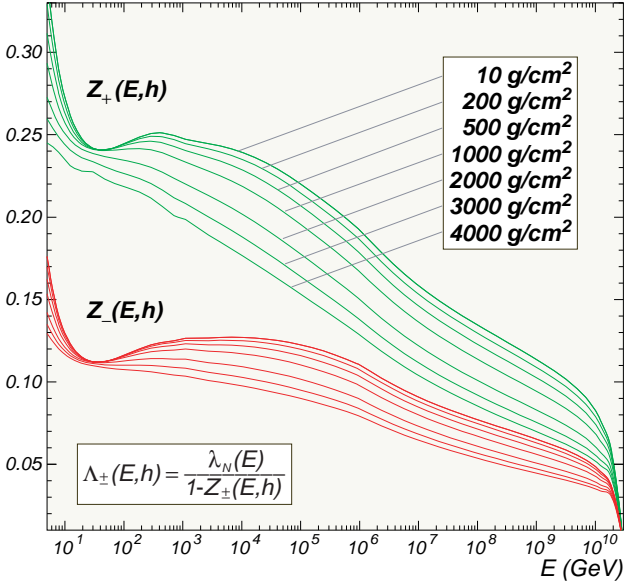} 
\fi
\protect\caption{$Z_+$ and $Z_-$ vs energy, computed with the EKS
                 primary spectrum for seven oblique atmospheric
                 depths. Both $Z_+$ and $Z_-$ decrease with
                 increasing depth.
\label{fig:Z}}
\end{figure}

Figure \ref{fig:Z} illustrates energy dependence of the $Z$
factors calculated using the EKS model of the primary spectrum for
several oblique depths. The ground of the observed dependence of
$Z_\pm$ on $E$ and $h$ is in the following three effects:
  (i) a non-power-law primary spectrum,
 (ii) energy dependence of the total inelastic cross section,
(iii) violation of Feynman scaling in $NA$ interactions.
As a result, the attenuation lengths are also energy and depth
dependent. This fact must be taken into account explicitly for
determining the nucleon interaction length $\lambda_N(E)$ from the
measured nucleon intensities in the atmosphere.
Local minima in the $Z$ factors that appear in the region around 45
GeV are due to the beginning of the growth of
$\sigma_{NA}^{\mathrm{inel}}(E)$. At not overly large depths, the
shape of the energy dependence visibly changes at $E\gtrsim10^6$ GeV,
which is caused by artificially introduced freezing of the growth of
the quasielastic peak in the reaction $pA \to pX$.
The vanishing of the $Z$ factors at $E=E_c$ is due to the supposed
cutoff of the primary spectrum at $E>E_c$.

\begin{figure}[t]
\vspace*{2.0mm} 
\ifpdf\includegraphics[width=8.3cm]{capr.pdf} 
\else \includegraphics[width=8.3cm]{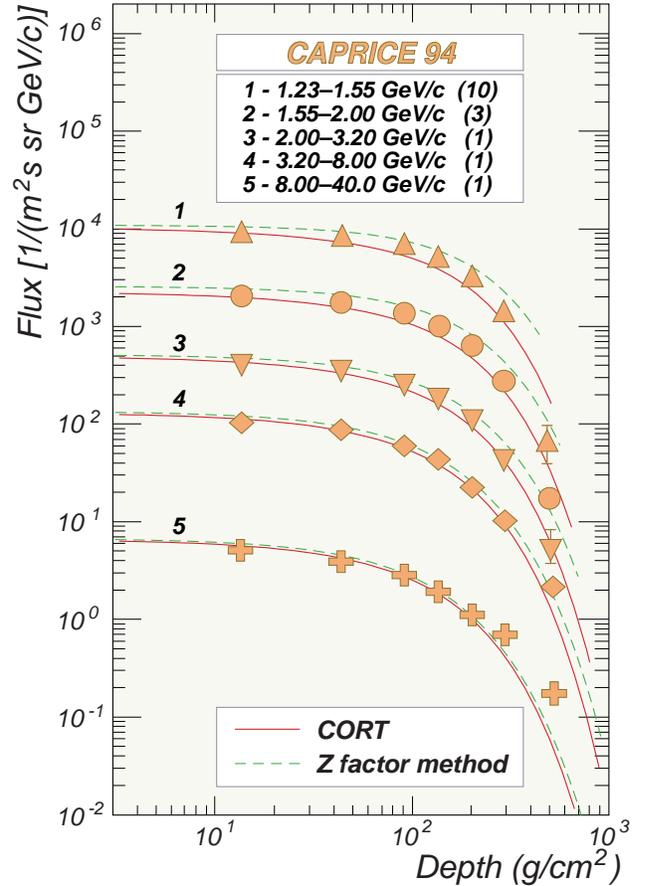} 
\fi
\protect\caption{Atmospheric growth curves for protons. The data
                 are from the CAPRICE\,94 balloon-borne experiment
                 \citep{Francke99}. The curves are calculated by
                 CORT and by using the $Z$ factor method with the
                 FNV model of primary spectrum. The legend shows
                 the proton momentum bins and scale factors
                 (in parentheses).
\label{fig:CAPRICE94}}
\end{figure}

Figure \ref{fig:CAPRICE94} shows the atmospheric growth of the
proton fluxes for five momentum bins. The data of the
balloon-borne experiment CAPRICE\,94 \citep{Francke99} are
compared with our calculations performed for the same bins by
using the $Z$ factor method and CORT code.
In both calculations we use the FNV model of the primary spectrum.
The nominal geomagnetic cutoff rigidity in the experiment was about
0.5 GV; hence the geomagnetic effect can be neglected.
As one can expect, the proton fluxes calculated by the
$Z$ factor method systematically exceed those are obtained with
CORT code, which takes into account the energy loss effect. The
discrepancies become larger at increasing depth but vanishes at
increasing the proton momentum. Minor differences between two
calculations at small depths are in part due to different
treatments of nucleus-nucleus interactions.

\begin{figure}[thb]
\vspace*{2.0mm} 
\ifpdf\includegraphics[width=8.3cm]{nucl.pdf} 
\else \includegraphics[width=8.3cm]{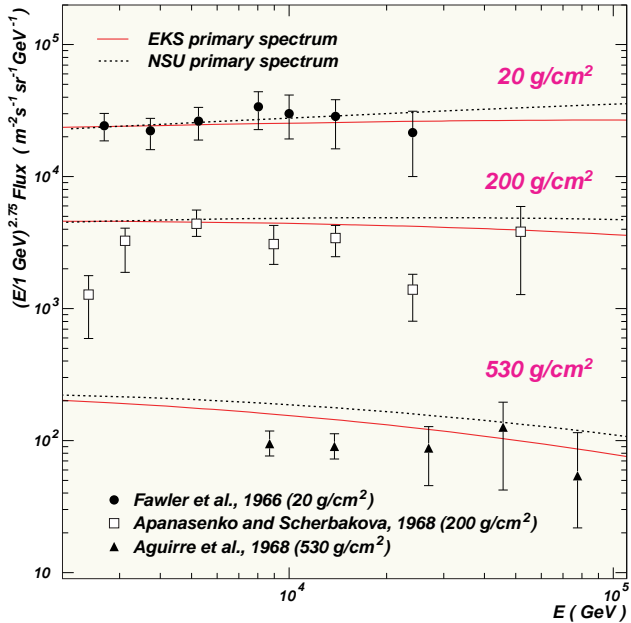} 
\fi
\protect\caption{Energy spectra of nucleons at three atmospheric
                 depths.
\label{fig:Nuc}}
\end{figure}

\begin{figure}[thb]
\vspace*{2.0mm} 
\ifpdf\includegraphics[width=8.3cm]{pn.pdf} 
\else \includegraphics[width=8.3cm]{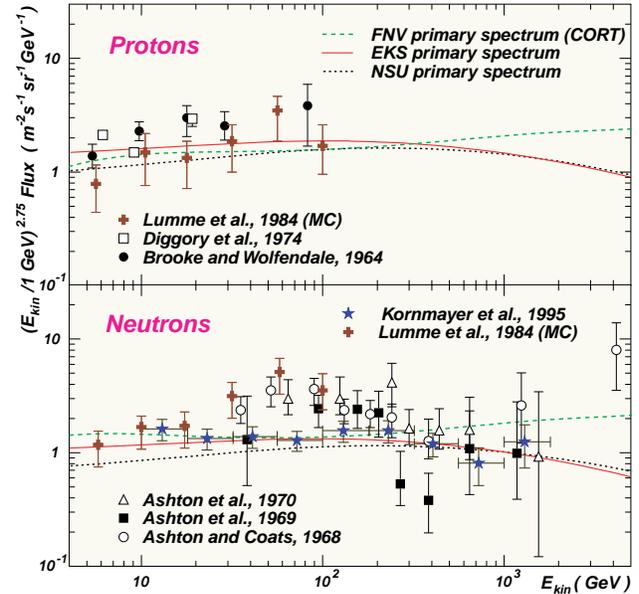} 
\fi
\protect\caption{Energy spectra of protons and neutrons at sea level.
\label{fig:P-N}}
\end{figure}


Comparison of the calculated differential energy spectra of
nucleons at three atmospheric depths with the data from
\citet{Fawler66,Aguirre68,Apanasenko68} are depicted in Fig.
\ref{fig:Nuc}. The data collected in Fig. \ref{fig:Nuc} were
obtained indirectly, from measurements of photon spectra in extensive
air showers \citep{Grigorov73} and are therefore model-dependent to
considerable extent.%
\footnote{In particular, approximate formulas used for recalculating
          photon energy to nucleon energy lead to a distortion of
          the nucleon energy spectra.}
Nonetheless, our calculation relying on the EKS and NSU models of
primary spectrum are by and large consistent with this data sample.

In Fig. \ref{fig:P-N} we compare the calculated energy spectra of
protons and neutrons at sea level with the data from
\citet{Brooke64,Ashton68,Ashton69,Ashton70,Diggory74,Kornmayer95}.
The results of Monte Carlo calculation by \citet{Lumme84} is also
shown. Direct measurements of the proton energy spectra at sea
level are very fragmentary and one can speak only about
qualitative agreement with the results of our calculations.
Estimates reveal (see also \citet{Vall86,Bugaev98}) that the
inclusion of processes of nucleon production in meson-nucleus
interactions can increase the vertical flux of nucleons at sea
level by no more than 10\% at $E=1$ TeV and by about 15\% at
$E=10$ TeV, but this increase is much smaller, in either case,
than the uncertainties in the $NA$ cross sections and in the
primary spectrum. Experimental data on the neutron component at
sea level are more detailed, but rather contradictory. The results
of our calculations are in quite a good agreement with the data
from recent measurements at prototype of the KASCADE facility in
Karlsruhe \citep{Kornmayer95}. As can be seen from Fig.
\ref{fig:P-N}, the neutron data by \citet{Kornmayer95} below
200--300 GeV are described by the calculation with the EKS and FNV
models of primary spectrum somewhat better than by the calculation
with the NSU model.

It can be hoped that further experiments to study the nucleon
component of secondary cosmic rays will allow a more detailed test of
the method and of the models for the primary spectrum and for
nucleon-nucleus interactions.

\begin{acknowledgements}
This work is supported by the Ministry of Education of the Russian
Federation under  Grant No. 015.02.01.004 (the Program
``Universities of Russia -- Basic Researches'').
\end{acknowledgements}

\end{document}